# EGYPTIAN RATSCREW: DISCOVERING DOMINANT STRATEGIES WITH COMPUTATIONAL GAME THEORY

Justin Diamond and Ben Garcia

---

A Thesis Presented to the Honors Program
of the George Washington University

---

Adviser: Professor Sumit Joshi
April 2022

## Abstract

"Egyptian Ratscrew" (ERS) is a modern American card game enjoyed by millions of players worldwide. A game of ERS is won by collecting all of the cards in the deck. Typically this game is won by the player with the fastest reflexes, since the most common strategy for collecting cards is being the first to slap the pile in the center whenever legal combinations of cards are placed down. Most players assume that the dominant strategy is to develop a faster reaction time than your opponents. Despite the fact that Egyptian Ratscrew has enjoyed decades of popularity, no academic inquiry has been levied against this assumption. This thesis investigates the hypothesis that a "risk slapping" strategist who relies on practical economic decision making will win an overwhelming majority of games against players who rely on quick reflexes alone. It is theorized that this can be done by exploiting the "burn rule," a penalty that is too low-cost to effectively dissuade players from slapping illegally when it benefits them.

Using the Ruby programming language, we construct an Egyptian Ratscrew simulator from scratch. Our modeling allows us to simulate the behavior of 8 strategically unique players within easily adjustable parameters including simulation type, player count, and burn amount.
We simulate 100k iterations of 67 different Egyptian Ratscrew games, totaling 6.7 million games of ERS, and use win percentage data in order to determine which strategies are dominant under each set of parameters. We then confirm our hypothesis that risk slapping is a dominant strategy, discover that there is no strictly dominant approach to risk slapping, and elucidate a deeper understanding of different ERS mechanics such as the burn rule. Finally, we assess the implications of our findings and suggest potential improvements to the rules of the game. We also touch on the real-world applications of our research and make recommendations for the future of Egyptian Ratscrew modeling.

## Acknowledgements

**Justin Diamond:**

The completion of this thesis would not be possible without the guidance of my adviser, Professor Sumit Joshi. I thank him for his time, advice, and encouragement this semester. I must also thank my co-author, Ben Garcia, for his incredible programming talent. Our modeling and simulations would be hollow and incomplete without his contributions.

I thank my family for supporting me throughout my time at university. Without our collective perfectionism, I'd have never held myself to such high standards. My Mom's meticulous approach, my Dad's work ethic, my Grandparents' unconditional love, my older sister Danielle's originality, and my younger sister Brooke's curiosity are hopefully all reflected in my research.

I'd also like to thank my roommates and lifelong friends, who've spent countless hours playing games with me, including ERS. My game theories could never be applied without Nick, Dan, Haining, Ian, and Ethan.

**Ben Garcia:**

Thanks to my coworkers, especially Eric and Maddy, for showing me just how bad I am at ERS. Their speed and ability to simply outclass me fueled my desire to work on this project.

Thanks to my family and girlfriend Olivia, for being the only people I could convince to listen to me talk about building an ERS simulator. The talking-it-through-out-loud phase is an often underappreciated step in software development. I could not have done it without them.

Lastly, thank you to Justin Diamond, principle Author, for throwing an absolute hail mary by reaching out to me at such a serendipitously fantastic time, the morning after I was embarrassed by Eric and Maddy in this exact game.

# Chapter 1:
# Understanding Egyptian Ratscrew

## 1.1 Introduction

"Egyptian Ratscrew" (ERS) is a modern American card game enjoyed by millions of players worldwide. The game goes by many names including but not limited to "Egyptian Ratslap," "Egyptian Rhapsody," "Egyptian War," and "Bloodystump." ERS likely evolved from the 19th-century British card game "Beggar-My-Neighbour," (BMN) which has strikingly similar rules. The main difference between these games is that ERS introduces elements of decision making, speed, and violence. Unlike BMN, whose winner is predetermined by the random order in which cards are dealt at the outset, games of ERS are traditionally won by the player with the fastest reflexes, in a skill-based manner.

Mathematicians have extensively studied the finiteness of BMN, primarily via brute-force computation. Thus far billions of "deals," or legal starting distributions of cards, have been analyzed. Still, no permutations have been found to result in a game which cycles infinitely.[1] Although Beggar-My-Neighbour is foundationally similar to the game we know and love today, we are not concerned with the mathematical finiteness of this antiquated and luck-based contest. Instead, we are concerned with its skill-based successor, Egyptian Ratscrew.

We believe that there is potential to completely and permanently reshape the way that millions of players approach Egyptian Ratscrew, especially given the complete lack of attention that has been paid toward discovering its hidden secrets and dominant strategies. By exploiting key features of the slapping apparatus, we hope that a new playing style based on practical economic decision making can undermine the supremacy of players who rely on quick reflexes or even card counting.

## 1.2 Rules of the Game

How are games of BMN and ERS played? It is important to note that every rule of BMN is *also* a rule of ERS, and ERS is only distinguished by its added concept of "slapping" cards. Below are the rules of BMN, as described by mathematician Marc M. Paulhus:

> At the start, an ordinary deck of 52 cards is divided as equally as possible among the players, who hold their respective shares face down. Players in rotation take one card from the top of their stack and place it face up on a stack in the center of the table. Play continues until a face card (J, Q, K, or A) is played, whereupon the next player is required to contribute respectively 1, 2, 3, or 4 cards to the central stack. If one of these 1, 2, 3, or 4 cards is a face card, then the player stops contributing, and the onus to supply the appropriate number of cards to the central stack passes to the next player. If none of the 1, 2, 3, or 4 cards is a face card, then the last player to play a face card collects the whole of the central stack, turns it over, and adds it to the underside of his own stack. This player then starts play again by turning the top card of his stack and placing it face up in the center of the table. Any player who runs out of cards, drops out of the game. If this happens during the contribution process, then the obligation to complete the contribution passes to the next player. The winner is the player who accumulates the whole deck.[2]

[1] Paulhus, Marc M. "Beggar My Neighbour." The American Mathematical Monthly 106, no. 2 (1999): 162–165.

When additional "slapping" rules are implemented, then a friendly game of BMN becomes a violent bout of ERS. In ERS, players gain the ability to acquire cards by slapping the central stack when certain combinations arise. Only the most conventional and widely accepted combinations will be considered in this study, and are listed below:

Doubles: Two like cards in a row. (2,2/3,3)

Sandwiches: A card sandwiched between two like cards. (2,3,2/5,4,5)

Tens: Two cards which equal 10 when added. (A,9/2,8/3,7/4,6)

Straights: Three cards in a row, ascending or descending. (2,3,4/K,Q,J)

Top-Bottom: When the card on top matches the card on the bottom. (2,7,K,4,9,2)

Marriage: When a king is next to a queen, or a queen next to a king. (K,Q/Q,K)

There are many different legal combinations, some of which develop regionally and others of which remain obscure house-rules. Alternative combos can include Jokers, Flushes, or Double Sandwiches – though there are countless more. When the deck is slapped in the absence of a proper combination, the next card in the player's deck is "burned" by placing it at the bottom of the central stack. This penalty is known as the burn rule, and it is supposed to dissuade players from rushing to slap, or slapping illegally. This "burned" card will become the new bottom card, in terms of applying the top-bottom rule. In games with >2-Players, those who are eliminated early will sometimes be allowed to slap back into the game. If an illegal slap is made by a player who is already out, then they are permanently eliminated. As in BMN, the winner of an Egyptian Ratscrew game is the player who manages to collect the whole deck of cards.

[2] Paulhus, Marc M. "Beggar My Neighbour." The American Mathematical Monthly 106, no. 2 (1999): 163.

# Chapter 2:

# Hypothesis, Modeling, and Methodology

## 2.1 Hypothesis

The slapping apparatus of Egyptian Ratscrew is what makes it so fast, violent, and distinct from other card collecting games of our time. At first glance, this mechanism doesn't seem to generate any particularly complex dominant strategies. Most players readily assume that the simplest approach is also the strongest one: Slap as quickly as possible, whenever legal combinations arise. Despite the fact that Egyptian Ratscrew has enjoyed decades of popularity, no academic inquiry has been levied against this assumption.

We hypothesize that under conventional rules of ERS, the dominant strategy is to begin preemptively slapping once the cards in the central stack meet a certain condition. This type of slap will be referred to as a "risk slap," and it must occur before the newly placed card can even be seen. Despite risking the low-cost penalty of burning a card, this strategy guarantees that if a legal combination arises, the risk slapper will win all of the cards in the central stack. This is much like committing a crime because the fine for getting caught is way lower than the expected payoff. We believe that by exploiting the burn rule and using well timed "risk slaps," a strategist who relies on practical economic decision making will win an overwhelming majority of games against players who rely on quick reflexes alone.

## 2.2 Modeling

In our quest to model Egyptian Ratscrew as precisely as possible, we opted for a computational and empirical approach. Only rigorous and repeatable simulations could appropriately model the near-infinite permutations of cards, combinations, and strategies. Using Ruby as our programming language, we coded from scratch an Egyptian Ratscrew simulator.[3] All rules of the game, as described in section 1.2, have been integrated into this simulator, with a few inconsequential exceptions:

- If a player holds no cards in their hand, risk slapping becomes disallowed. This prevents players from slapping illegally when they have no cards left to burn.
- Slapping the last card in a sequence following a face card is disallowed. This is inconsequential because any player who places a face card will always slap at the final card in such a sequence, as they are entitled to collect the deck even when there is no legal slap.

[3] Garcia, Ben. "Benngarcia/Egyptian-Ratscrew-Ruby." GitHub, April 22, 2022.

- Slapping back into the game will not be allowed for or considered within this study. In games with >2-Players, elimination from the game is permanent.

The model is also dynamic, and allows users to run tests under different conditions. Strategist type, simulation type, number of game iterations, player count, and burn amount are all easily adjustable parameters.

#### 2.2.1 Strategist Types

In our simulations, there are 3 types of players and 8 unique strategies. These players and strategies are characterized as follows:

1. Reflexive Players: Reflexive players will slap as quickly as possible, but only when legal combinations arise. We will not refer to Reflexives as "strategists," since they only execute the most rudimentary strategy. Abbreviated as "Ref"
2. Qualitative Strategists: Qualitative strategists will risk slap whenever face cards, or particular face cards, are contained in the central stack. This strategist values only face cards, which force opponents to contribute more of their deck to the central pile when played. Abbreviated as "Qual"
    - 2.1. Qual All Faces: Risk slaps whenever a face card is contained in the central stack, only after it has been placed. Abbreviated as "Qual All"
    - 2.2. Qual Jack through King: Risk slaps whenever a J, Q, or K is contained in the central stack, only after it has been placed. Abbreviated as "Qual J-K"
3. Quantitative Strategists: Quantitative strategists will risk slap whenever a certain number of cards are contained in the central stack. This strategist values any pile of cards, as long as it contains $n$ number of cards within it. Abbreviated as "Quant"
    - 3.1. Quant n=2: Risk slaps whenever 2 cards are contained in the central stack, including when the 2nd card is placed (and so on for other Quants).
    - 3.2. Quant n=3: Risk slaps whenever 3 cards are contained in the central stack.
    - 3.3. Quant n=4: Risk slaps whenever 4 cards are contained in the central stack.
    - 3.4. Quant n=5: Risk slaps whenever 5 cards are contained in the central stack.
    - 3.5. Quant n=6: Risk slaps whenever 6 cards are contained in the central stack.

Our model does not include strategies like card counting. Although card counting can be incredibly effective in ERS, it is too difficult for most humans to employ in real time – especially given the speed at which ERS is played. We also ignore the implementation of mixed strategies, but hope that they will be studied comprehensively in the future.

### 2.2.2 Simulation Types

All simulations within our model have a particular "strategic speed." Strategic speed can be interpreted as the specialization in, or success rate of, implementing one's own strategy. For example, in a game with 80% strategic speed, a risk slapper will have an 80% chance of winning risk slaps and a Reflexive player will have an 80% chance of winning speed slaps. The opponent(s) in each case will have the 20% chance of winning that exists as a remainder. We implemented this feature in order to represent imperfect play, and reflect more accurately the human error in executing any given strategy. Strategic speed exists on an easily adjustable sliding scale from 0–100%, but in broad terms we can characterize our simulations as existing in two categories:

1. Controlled Simulations: These simulations reflect the efficacy of each strategy in its purest form, and assumes that players will completely specialize in their respective strategies. Controlled simulations are run strictly at 100% strategic speed, such that risk slappers will win 100% of hands when risk slapping and Reflexive players will win 100% of hands when speed slapping.
2. Probabilistic Simulations: These simulations reflect the efficacy of each strategy when executed imperfectly and without complete specialization. Probabilistic simulations can be run at any strategic speed below 100%, but we can reasonably guess that the most "human" degree of strategic specialization lies somewhere around 80 or 90%.

### 2.2.3 Imperfections of the Model

As dynamic and precise as our simulator might be, there remain some imperfections within our model. Three main shortcomings have been identified, two of which have been solved or pseudo-solved.

The first and least consequential imperfection was that when running large numbers of multi-strategist games, an odd situation occured wherein both players had burned 100% of their cards without a winner being selected. We solved this problem by having the simulator randomly select a winner. Although this exceptionally rare circumstance is of little interest, it felt like our model was paying homage to the many mathematicians before us who studied the finiteness of Beggar-My-Neighbour. In a way, we had stumbled upon a somewhat-infinite variation of the game in which no winner could be determined.

The second imperfection was that in some types of games, the ordering of strategists was affecting our results. This problem was pseudo-solved by shuffling the order of strategists prior to every iteration of a game.

The third, and most consequential imperfection is the quasi-team problem. This problem only occurs in games with 2 or more players of a common strategy. In these games, our model treats players with common strategies as if they are on a quasi-team. Players who share a strategy will split the share of hands that are won by said strategy in that game. This leads to a slight overperformance of lone strategists in games with 2 or more players of a common strategy.

#### 2.2.4 Open Source Code

The simulator used for all of our modeling is freely available in the form of open source code.[4] All of our results are easily reproducible, and we hope that others will run more tests while discovering new strategies and improving upon our model of Egyptian Ratscrew. The documentation provided should help others make use of this simulator.

### 2.3 Methodology

We will simulate 100k iterations of 67 different Egyptian Ratscrew games, totaling 6.7 million games of ERS. We will rely solely on win percentage – the most straightforward metric of success – in order to determine which strategies work best under different parameters. Below is an exhaustive list of every game that will be simulated:

2-Player Games:

1. 100k Qual All v Ref, 100% Strategic Speed
2. 100k Qual All v Ref, 90% Strategic Speed
3. 100k Qual All v Ref, 80% Strategic Speed
4. 100k Qual All v Ref, 75% Strategic Speed
5. 100k Qual All v Ref, 70% Strategic Speed
6. 100k Qual All v Ref, 60% Strategic Speed
7. 100k Qual All v Ref, 50% Strategic Speed
8. 100k Qual J-K v Ref, 100% Strategic Speed
9. 100k Qual J-K v Ref, 90% Strategic Speed
10. 100k Qual J-K v Ref, 80% Strategic Speed
11. 100k Qual J-K v Ref, 75% Strategic Speed
12. 100k Qual J-K v Ref, 70% Strategic Speed
13. 100k Qual J-K v Ref, 60% Strategic Speed
14. 100k Qual J-K v Ref, 50% Strategic Speed
15. 100k Quant n=2 v Ref, 100% Strategic Speed
16. 100k Quant n=3 v Ref, 100% Strategic Speed
17. 100k Quant n=3 v Ref, 90% Strategic Speed

[4] Garcia, Ben. "Benngarcia/Egyptian-Ratscrew-Ruby." GitHub, April 22, 2022.

18. 100k Quant n=3 v Ref, 80% Strategic Speed
19. 100k Quant n=3 v Ref, 75% Strategic Speed
20. 100k Quant n=3 v Ref, 70% Strategic Speed
21. 100k Quant n=3 v Ref, 60% Strategic Speed
22. 100k Quant n=3 v Ref, 50% Strategic Speed
23. 100k Quant n=4 v Ref, 100% Strategic Speed
24. 100k Quant n=5 v Ref, 100% Strategic Speed
25. 100k Quant n=6 v Ref, 100% Strategic Speed

4-Player Games:

26. 100k Qual All v Ref (x3), 100% Strategic Speed
27. 100k Qual All v Ref (x3), 90% Strategic Speed
28. 100k Qual All v Ref (x3), 80% Strategic Speed
29. 100k Qual All v Ref (x3), 75% Strategic Speed
30. 100k Qual All v Ref (x3), 70% Strategic Speed
31. 100k Qual All v Ref (x3), 60% Strategic Speed
32. 100k Qual All v Ref (x3), 50% Strategic Speed
33. 100k Qual J-K v Ref (x3), 100% Strategic Speed
34. 100k Qual J-K v Ref (x3), 90% Strategic Speed
35. 100k Qual J-K v Ref (x3), 80% Strategic Speed
36. 100k Qual J-K v Ref (x3), 75% Strategic Speed
37. 100k Qual J-K v Ref (x3), 70% Strategic Speed
38. 100k Qual J-K v Ref (x3), 60% Strategic Speed
39. 100k Qual J-K v Ref (x3), 50% Strategic Speed
40. 100k Quant n=2 v Ref (x3), 100% Strategic Speed
41. 100k Quant n=3 v Ref (x3), 100% Strategic Speed
42. 100k Quant n=3 v Ref (x3), 90% Strategic Speed
43. 100k Quant n=3 v Ref (x3), 80% Strategic Speed
44. 100k Quant n=3 v Ref (x3), 75% Strategic Speed
45. 100k Quant n=3 v Ref (x3), 70% Strategic Speed
46. 100k Quant n=3 v Ref (x3), 60% Strategic Speed
47. 100k Quant n=3 v Ref (x3), 50% Strategic Speed
48. 100k Quant n=4 v Ref (x3), 100% Strategic Speed
49. 100k Quant n=5 v Ref (x3), 100% Strategic Speed
50. 100k Quant n=6 v Ref (x3), 100% Strategic Speed

Multi-Strategist Games:

51. 100k Qual J-K v Qual All, 100% Strategic Speed
52. 100k Qual J-K v Quant n=3, 100% Strategic Speed
53. 100k Qual All v Quant n=3, 100% Strategic Speed
54. 100k Qual J-K v Qual All v Quant n=3 v Reflexive, 100% Strategic Speed

Supplemental Games:

55. 100k Qual All v Ref (x7), 8-Player, 100% Strategic Speed
56. 100k Qual All v Ref (x7), 8-Player, 90% Strategic Speed
57. 100k Qual All v Ref (x15), 16-Player, 100% Strategic Speed
58. 100k Qual All v Ref (x15), 16-Player, 90% Strategic Speed
59. 100k Qual All v Ref, 100% Strategic Speed, Burn amount 0

60. 100k Qual All v Ref, 100% Strategic Speed, Burn amount 2
61. 100k Qual All v Ref, 100% Strategic Speed, Burn amount 3
62. 100k Qual All v Ref, 100% Strategic Speed, Burn amount 4
63. 100k Qual All v Ref, 100% Strategic Speed, Burn amount 5
64. 100k Qual All v Ref, 90% Strategic Speed, Burn amount 2
65. 100k Qual All v Ref, 90% Strategic Speed, Burn amount 3
66. 100k Qual All v Ref, 90% Strategic Speed, Burn amount 4
67. 100k Qual All v Ref, 90% Strategic Speed, Burn amount 5

The bulk of these tests should provide us with a clear idea of how powerful each strategy is, which strategy is most dominant, and whether or not there is a strictly dominant strategy that remains effective regardless of an opponents' ability to change behavior. In addition, supplemental tests will help elucidate a deeper understanding of the game's mechanics by modifying key parameters such as player count and burn amount.

# Chapter 3:

# Results and Findings

### 3.1 Comprehensive Simulation Results

*Figure 1: Comprehensive results of all 6.7 million simulated Egyptian Ratscrew games*

| 2-Player Games: | Win Rate | Win Rate | Win Rate | Win Rate |
|---|---|---|---|---|
| 100k Qual All v Ref, 100% | 90.689% | 9.311% | | |
| 100k Qual All v Ref, 90% | 80.229% | 19.771% | | |
| 100k Qual All v Ref, 80% | 61.750% | 38.250% | | |
| 100k Qual All v Ref, 75% | 49.223% | 50.777% | | |
| 100k Qual All v Ref, 70% | 36.823% | 63.177% | | |
| 100k Qual All v Ref, 60% | 16.175% | 83.825% | | |
| 100k Qual All v Ref, 50% | 5.720% | 94.280% | | |
| 100k Qual J-K v Ref, 100% | 89.418% | 10.582% | | |
| 100k Qual J-K v Ref, 90% | 79.017% | 20.983% | | |
| 100k Qual J-K v Ref, 80% | 60.196% | 39.804% | | |
| 100k Qual J-K v Ref, 75% | 47.811% | 52.189% | | |
| 100k Qual J-K v Ref, 70% | 35.448% | 64.552% | | |
| 100k Qual J-K v Ref, 60% | 15.759% | 84.241% | | |
| 100k Qual J-K v Ref, 50% | 5.705% | 94.295% | | |
| 100k Quant n=2 v Ref, 100% | 82.242% | 17.758% | | |
| 100k Quant n=3 v Ref, 100% | 82.994% | 17.006% | | |

| 100k Quant n=3 v Ref, 90% | 65.845% | 34.155% | | |
|---|---|---|---|---|
| 100k Quant n=3 v Ref, 80% | 42.303% | 57.697% | | |
| 100k Quant n=3 v Ref, 75% | 30.531% | 69.469% | | |
| 100k Quant n=3 v Ref, 70% | 21.188% | 78.812% | | |
| 100k Quant n=3 v Ref, 60% | 8.513% | 91.487% | | |
| 100k Quant n=3 v Ref, 50% | 2.944% | 97.056% | | |
| 100k Quant n=4 v Ref, 100% | 65.288% | 34.712% | | |
| 100k Quant n=5 v Ref, 100% | 19.349% | 80.651% | | |
| 100k Quant n=6 v Ref, 100% | 3.438% | 96.562% | | |
| 4-Player Games: | Win Rate | Win Rate | Win Rate | Win Rate |
| 100k Qual All v Ref (x3), 100% | 73.118% | 26.882% | | |
| 100k Qual All v Ref (x3), 90% | 59.992% | 40.008% | | |
| 100k Qual All v Ref (x3), 80% | 42.182% | 57.818% | | |
| 100k Qual All v Ref (x3), 75% | 31.959% | 68.041% | | |
| 100k Qual All v Ref (x3), 70% | 22.649% | 77.351% | | |
| 100k Qual All v Ref (x3), 60% | 8.545% | 91.455% | | |
| 100k Qual All v Ref (x3), 50% | 2.418% | 97.582% | | |
| 100k Qual J-K v Ref (x3), 100% | 71.468% | 28.532% | | |
| 100k Qual J-K v Ref (x3), 90% | 59.192% | 40.808% | | |
| 100k Qual J-K v Ref (x3), 80% | 42.396% | 57.604% | | |
| 100k Qual J-K v Ref (x3), 75% | 32.364% | 67.636% | | |
| 100k Qual J-K v Ref (x3), 70% | 23.337% | 76.663% | | |
| 100k Qual J-K v Ref (x3), 60% | 9.428% | 90.572% | | |
| 100k Qual J-K v Ref (x3), 50% | 3.027% | 96.973% | | |
| 100k Quant n=2 v Ref (x3), 100% | 65.986% | 34.014% | | |
| 100k Quant n=3 v Ref (x3), 100% | 70.958% | 29.042% | | |
| 100k Quant n=3 v Ref (x3), 90% | 53.546% | 46.454% | | |
| 100k Quant n=3 v Ref (x3), 80% | 31.939% | 68.061% | | |
| 100k Quant n=3 v Ref (x3), 75% | 22.180% | 77.820% | | |
| 100k Quant n=3 v Ref (x3), 70% | 14.100% | 85.900% | | |
| 100k Quant n=3 v Ref (x3), 60% | 4.807% | 95.193% | | |
| 100k Quant n=3 v Ref (x3), 50% | 1.284% | 98.716% | | |
| 100k Quant n=4 v Ref (x3), 100% | 58.703% | 41.297% | | |
| 100k Quant n=5 v Ref (x3), 100% | 18.387% | 81.613% | | |
| 100k Quant n=6 v Ref (x3), 100% | 2.743% | 97.257% | | |

| Multi-Strategist Games: | Win Rate | Win Rate | Win Rate | Win Rate |
| --- | --- | --- | --- | --- |
| 100k Qual J-K v Qual All, 100% | 50.832% | 49.168% | | |
| 100k Qual J-K v Quant n=3, 100% | 61.774% | 38.226% | | |
| 100k Qual All v Quant n=3, 100% | 58.784% | 41.216% | | |
| 100k Qual J-K v Qual All v Quant n=3 v Ref, 100% | 33.403% | 31.506% | 26.880% | 8.211% |
| Supplemental Games: | Win Rate | Win Rate | Win Rate | Win Rate |
| 100k Qual All v Ref (x7), 8-Player, 100% | 53.270% | 46.730% | | |
| 100k Qual All v Ref (x7), 8-Player, 90% | 41.180% | 58.820% | | |
| 100k Qual All v Ref (x15), 16-Player, 100% | 35.616% | 64.384% | | |
| 100k Qual All v Ref (x15), 16-Player, 90% | 26.676% | 73.324% | | |
| 100k Qual All v Ref, 100%, Burn amount 0 | 99.874% | 0.126% | | |
| 100k Qual All v Ref, 100%, Burn amount 2 | 62.995% | 37.005% | | |
| 100k Qual All v Ref, 100%, Burn amount 3 | 40.676% | 59.324% | | |
| 100k Qual All v Ref, 100%, Burn amount 4 | 27.098% | 72.902% | | |
| 100k Qual All v Ref, 100%, Burn amount 5 | 18.895% | 81.105% | | |
| 100k Qual All v Ref, 90%, Burn amount 2 | 45.946% | 54.054% | | |
| 100k Qual All v Ref, 90%, Burn amount 3 | 26.787% | 73.213% | | |
| 100k Qual All v Ref, 90%, Burn amount 4 | 17.089% | 82.911% | | |
| 100k Qual All v Ref, 90%, Burn amount 5 | 11.725% | 88.275% | | |

### 3.2 2-Player Games

Figures 2 and 3 (next page) show that in 2-Player Games against Reflexive players, at most strategic speeds, Qual All is the most dominant strategy. The Qual J-K strategy is almost as dominant, but remains marginally weaker at all strategic speeds. In controlled experiments with 100% strategic speed, Qual All, Qual J-K, Quant n=2, Quant n=3, and Quant n=4 all manage to win an overwhelming majority of games against Reflexive players. Even at 80% strategic speed, Qualitative strategists continue to win over 60% of games. This shows that sub-optimal approaches to risk slapping and/or imperfect execution of the strategy still result in consistent wins against players who rely solely on reflexes. With perfect play, a Qual All strategist will win over 90% of games against a Reflexive player. This convincingly confirms the hypothesis that under conventional rules of ERS, risk slapping is a dominant strategy.

As dominant as these risk slapping strategies might be, they also leave little room for error. Since risk slappers are committed to burning many cards, they must win a certain number of central stacks in order to justify the high strategic cost. At lower strategic speeds, the win rate falls dramatically. This reflects the fact that risk slapping fundamentally relies on a high-risk, high-reward behavior that is supposed to pay off in the long run. At 50% strategic speed, risk slappers win so few games because they are slapping at many illegal combinations, yet only win half of the legal combinations that arise. Fortunately, it is relatively easy for risk slappers to maintain a high strategic speed. This is because risk slappers preemptively decide to slap before other players can even observe the most recently placed card – a time-consuming behavior that Reflexive players rely on in order to determine whether or not a legal combination has arisen.

*Figure 2: 2-Player Game v Reflexive Matrix, Strategists' Win Rates*

| Strategic Speed | 2-Player Games | Qual All | Qual J-K | Quant n=2 | Quant n=3 | Quant n=4 | Quant n=5 | Quant n=6 |
|---|---|---|---|---|---|---|---|---|
| 100% | Reflexive | 90.689% | 89.418% | 82.242% | 82.994% | 65.288% | 19.349% | 3.438% |
| 90% | Reflexive | 80.229% | 79.017% | / | 65.845% | / | / | / |
| 80% | Reflexive | 61.750% | 60.196% | / | 42.303% | / | / | / |
| 75% | Reflexive | 49.223% | 47.811% | / | 30.531% | / | / | / |
| 70% | Reflexive | 36.823% | 35.448% | / | 21.188% | / | / | / |
| 60% | Reflexive | 16.175% | 15.759% | / | 8.513% | / | / | / |
| 50% | Reflexive | 5.720% | 5.705% | / | 2.944% | / | / | / |

*Figure 3: Comparing Optimized Qualitative and Quantitative Strategists*

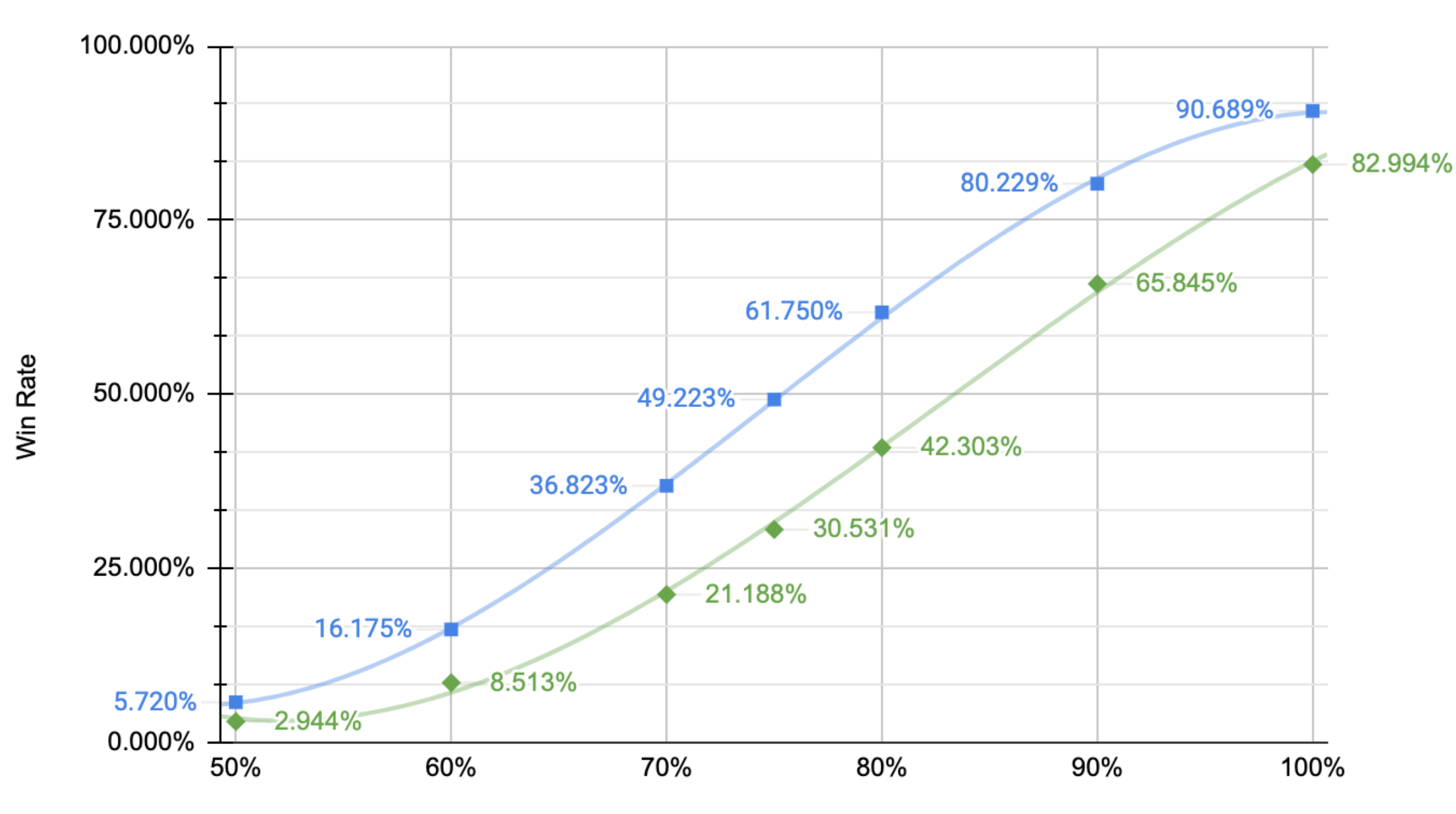

### 3.3 4-Player Games

Figure 4 shows that in 4-Player Games against Reflexive players, Qual All is the most dominant strategy at high strategic speeds. With perfect play, a Qual All strategist will win over 70% of games against 3 Reflexive players combined. Interestingly, the Qual J-K strategist manages to outperform the Qual All strategist at lower strategic speeds. However, this occurs only at speeds which result in a win rate below 50% – a point at which neither strategy is dominant. In controlled 4-Player experiments with 100% strategic speed, Qual All, Qual J-K, Quant n=2, Quant n=3, and Quant n=4 strategists manage to win a majority of games against Reflexives, with Qual All, Qual J-K, and Quant n=3 strategists winning >70% of these games. At 90% strategic speed, Qual All, Qual J-K, and Quant n=3 strategists continue to win >50% of games. Figure 5 (next page) shows just how impressive this is, since the theoretical win rate of a Reflexive player against 3 other Reflexive players is 25%. All this evidence further supports the hypothesis that risk slapping is a dominant strategy, by showing that it remains dominant in games with larger numbers of players.

Despite the risk slapping strategists' continued success in 4-Player games, we must reiterate the presence of an inherent flaw in our model. The quasi-team problem skews the results of all games in which 2 or more players share a common strategy, as is mentioned in section 2.2.3. These tests are biased in favor of lone strategists to an unknown degree, though we do not believe this distortion is consequential enough to alter general conclusions about which strategies are dominant in games with quasi-teams.

*Figure 4: 4-Player Game v Reflexive (x3) Matrix, Strategists' Win Rates*

| Strategic Speed | 4-Player Games | Qual All | Qual J-K | Quant n=2 | Quant n=3 | Quant n=4 | Quant n=5 | Quant n=6 |
|---|---|---|---|---|---|---|---|---|
| 100% | Reflexive (x3) | 73.118% | 71.468% | 65.986% | 70.958% | 58.703% | 18.387% | 2.743% |
| 90% | Reflexive (x3) | 59.992% | 59.192% | / | 53.546% | / | / | / |
| 80% | Reflexive (x3) | 42.182% | 42.396% | / | 31.939% | / | / | / |
| 75% | Reflexive (x3) | 31.959% | 32.364% | / | 22.180% | / | / | / |
| 70% | Reflexive (x3) | 22.649% | 23.337% | / | 14.100% | / | / | / |
| 60% | Reflexive (x3) | 8.545% | 9.428% | / | 4.807% | / | / | / |
| 50% | Reflexive (x3) | 2.418% | 3.027% | / | 1.284% | / | / | / |

*Figure 5: Comparing the Dominant Strategy in 2 and 4 Player Games*

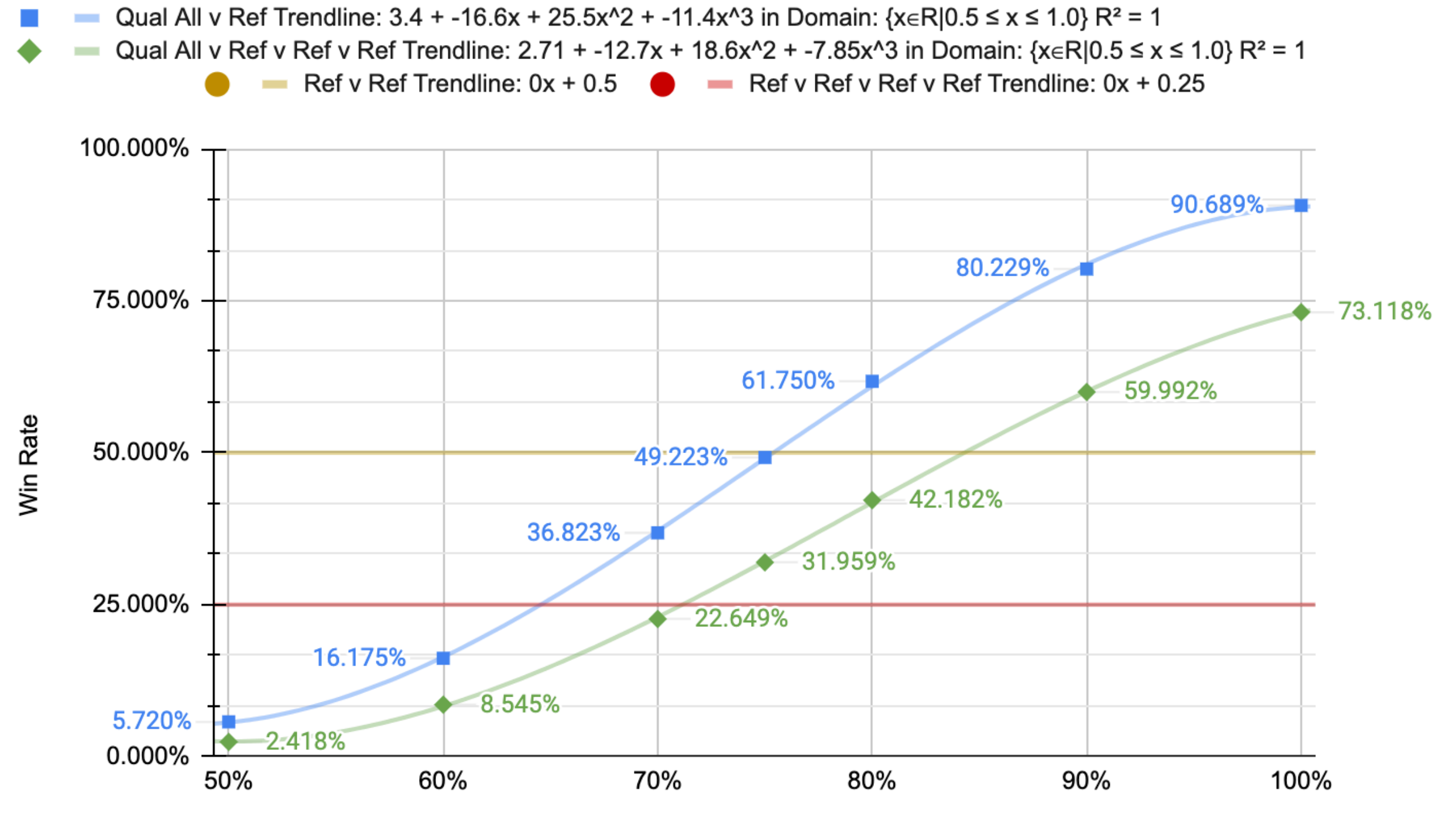

### 3.4 Multi-Strategist Games

Despite Qual All strategists' stronger performance against Reflexives in 2 and 4-Player games, Figure 6 shows that in Qual J-K v Qual All games, the Qual J-K strategist maintained a slight (<1%) advantage after 100,000 simulations. The Qual J-K strategist also beat the Quant n=3 strategist by wider margins than the Qual All strategist, achieving ~3% greater success. This evidence suggests that while risk slapping is generally the dominant strategy, in Egyptian Ratscrew there is no strictly dominant approach to risk slapping. In order to fully optimize one's strategy, a player must adapt to the changing behavior and strategy of their opponent(s).

*Figure 6: Multi-Strategist Game Results*

| Multi-Strategist Games: | Win Rate | Win Rate | Win Rate | Win Rate |
|---|---|---|---|---|
| 100k Qual J-K v Qual All, 100% | 50.832% | 49.168% | | |
| 100k Qual J-K v Quant n=3, 100% | 61.774% | 38.226% | | |
| 100k Qual All v Quant n=3, 100% | 58.784% | 41.216% | | |
| 100k Qual J-K v Qual All v Quant n=3 v Ref, 100% | 33.403% | 31.506% | 26.880% | 8.211% |

**3.5 Supplemental Games**

Figure 7 depicts the win rate of Qual All strategists in both controlled and probabilistic games, as the number of Reflexive opponents increases. We estimate that this data would exhibit a more exponential pattern, as is present in the theoretically derived Ref v Ref curve, if not for the quasi-team problem. This supplemental data is still useful because it reflects the increased risk of strategists burning out as more players are added to the game. For example, in a game with 16 players, each player starts with only 3 or 4 cards. For a risk slapper, this significantly increases the chance that they will burn all of their cards and be eliminated from the game before a legal combination can even be won.

*Figure 7: The Effects of Increasing Player Count on the Dominant Strategy*

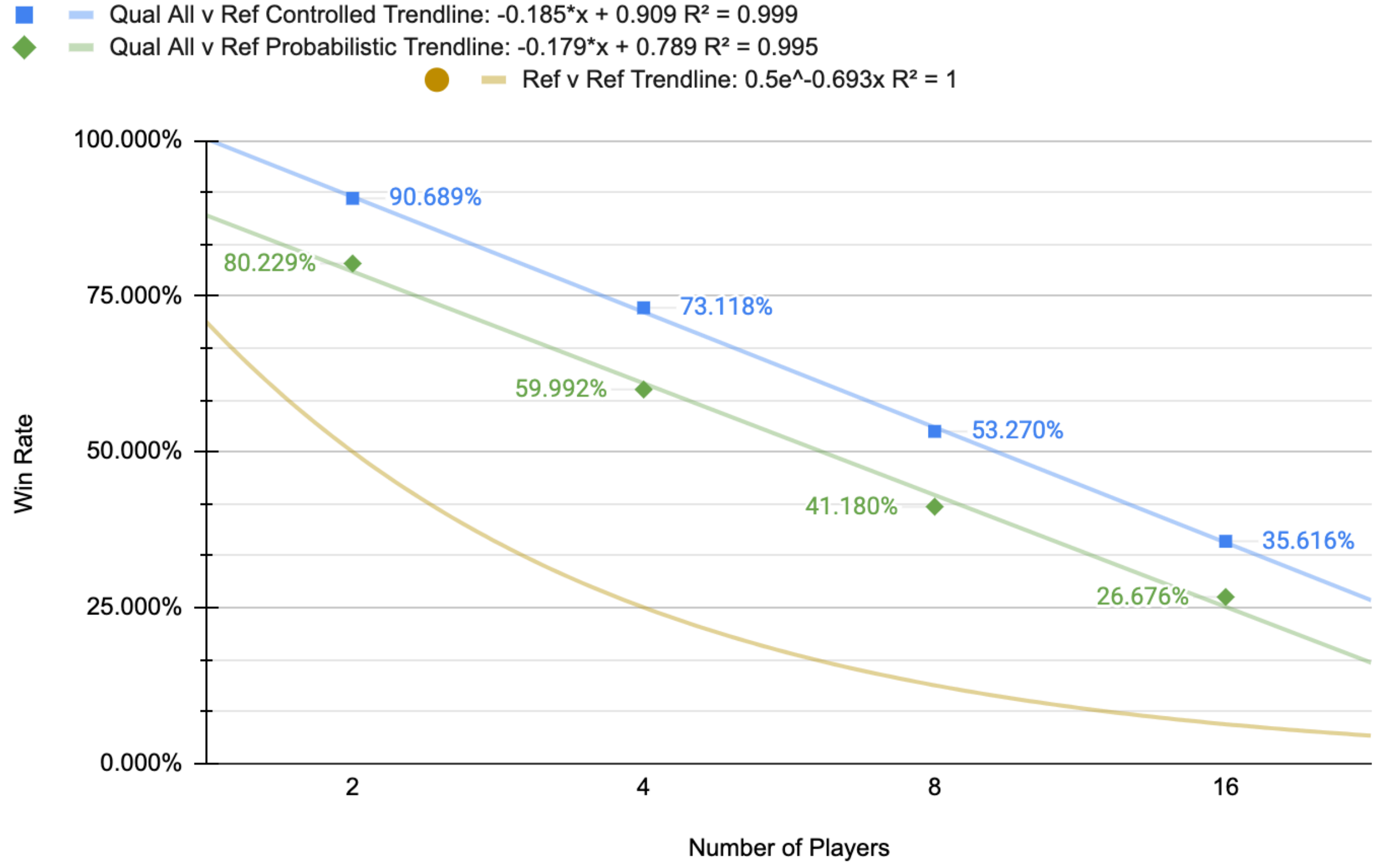

Figure 8 (next page) illustrates the effect of increasing burn penalties on the most dominant strategy's win rate. By looking at the effect of different penalties on the efficacy of the Qual All strategy in both controlled and probabilistic games, we can establish how large the penalty would have to become before risk slapping becomes infeasible. The data in Figure 8 shows that a 2 card penalty is sufficient to reduce the dominance of risk slapping such that Qual All strategists can only win a majority of games with perfect or near perfect play. Additionally, a penalty of 3 or more cards will completely eliminate the dominance of this particular risk slapping strategy – even with perfect play. This indicates that the expected value of the average Qual All risk slap becomes negative once the burn penalty is equal to or greater than 3 cards.

*Figure 8: The Effects of Larger Burn Penalties on the Dominant Strategy*

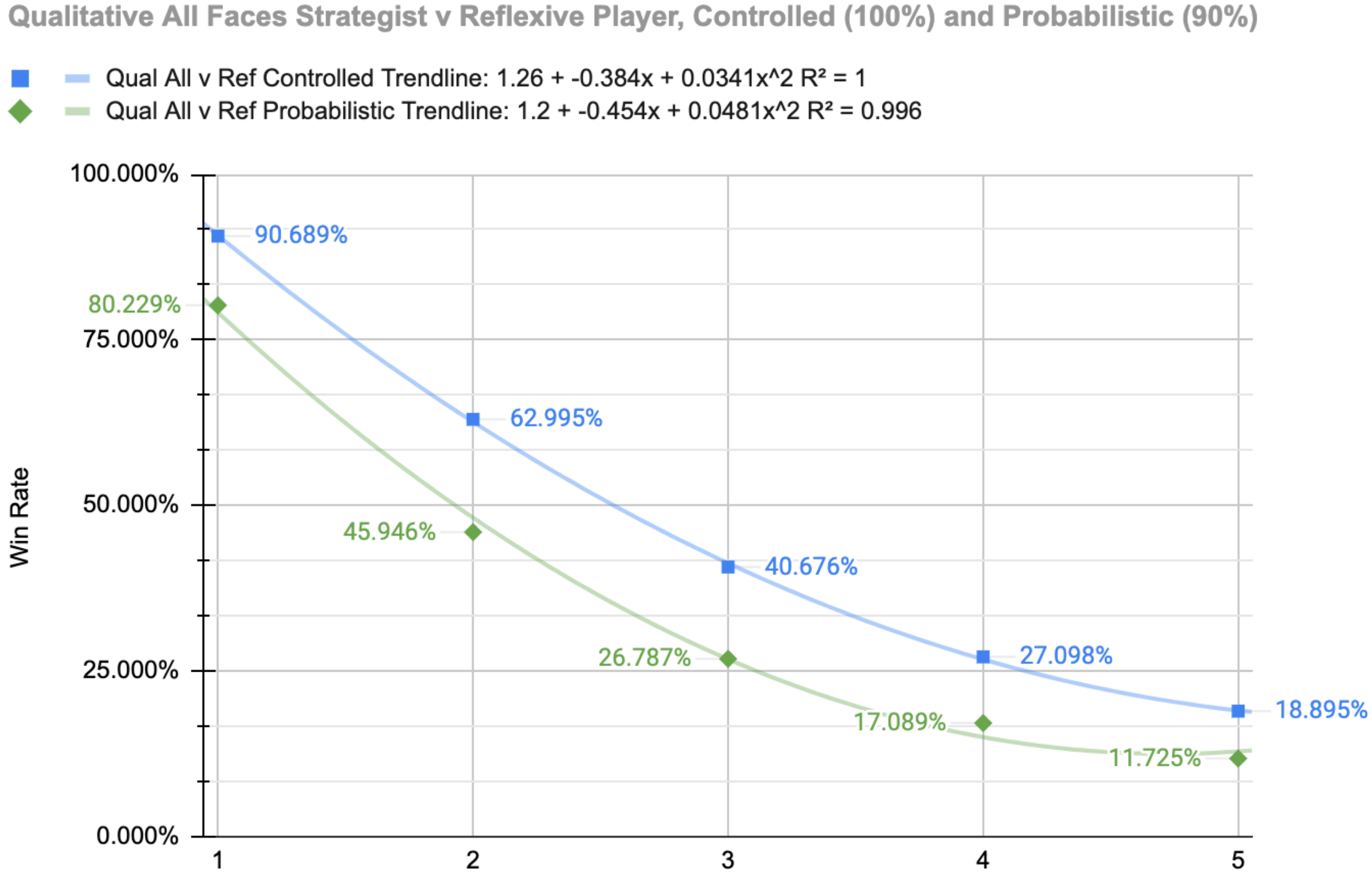

It is worth noting that many theoretical conditions for risk slapping have gone unexplored in this study. We have studied 7 unique risk slapping strategies, though countless variations upon them can be imagined. Alternative strategies such as slapping only when two face cards are contained in the deck may continue to be effective even after the burn penalty has been increased. This is because the central stack must contain more value before risk slapping begins. Nevertheless, these “large-penalty-proof” risk slappers will only slap in conditions that inherently occur less often, so we believe they can only succeed by implementing mixed strategies that rely to some extent upon fast reflexes.

# Chapter 4:

# Conclusions

### 4.1 Implications

Our results have confirmed the hypothesis that under conventional rules of ERS, risk slapping is a dominant strategy. This is especially true when executed by Qualitative strategists. Moreover, our simulations have dealt a fatal blow to purely Reflexive strategies. It is clear that while reflexes are a valuable attribute for any ERS player, speed slapping can not succeed on its own; players with quick reflexes must integrate their talents into a mixed-strategy approach. Another important implication is that there is no strictly dominant approach to risk slapping. The multi-strategist games in section 3.4 show us that identifying the best strategy in ERS is like hitting a constantly moving target; a strategy that is dominant against one type of player is not necessarily dominant against another. Lastly, if risk slapping happens too often, then ERS strategy may evolve to include countermeasures like feints. A feint is when a player quickly moves their card toward the center pile but hesitates before placing it down, provoking an illegal risk slap.

### 4.2 Rule Changes

We have omitted a uniquely philosophical implication from section 4.1, because it warrants having its own section for discussion. This is the meta-implication that our results will encourage players to exploit a flaw in the rules of this game, rather than improving its rules and conventions. Many players would see risk slapping as contrary to the spirit of Egyptian Ratscrew, since ERS is supposed to be a game of reflexes, speed, and even violence. If every player began regularly risk slapping, then most ERS games would quickly degenerate into gameplay that is somewhat hollow.

That being said, we want to suggest two new, equally enjoyable sets of rules. Both are meant to be implemented *ceteris paribus*, and their estimated effect on gameplay is substantiated by the data in Figure 8 of section 3.5:

1. 2 Card Burn Penalty: Under this ruleset, the dominant strategy is ambiguous, dynamic, and exciting. Risk slapping can work, but only if executed carefully.
2. 4 Card Burn Penalty: Under this ruleset, risk slapping is almost never profitable. This is for players who want to reward reflexes and speed only.

Of course, we prefer that all readers make use of our findings by beating their friends and loved ones senseless before implementing any changes to the conventional rules.

### 4.3 Real-World Applications

The dominance of risk slapping in ERS illustrates an economic principle that is widely applicable in the real-world. This principle is simple: Violating any given rule is beneficial to the rule breaker when the punitive cost is lower than the benefit gained by violating it. There are countless real-world examples of rules that can be broken profitably. For example, parking illegally in a town with low illegal-parking fines and expensive parking garages is an easy way to save money. Any rule with explicitly defined punitive costs can be put through the same kind of cost-benefit analysis that we have completed with regard to ERS.

The crude solution to unreasonably low punitive costs is to raise punitive costs, but in many real-world situations this can not always be done. In the realm of international affairs, many high-stakes dilemmas become difficult to address given the challenge of enforcing large punitive costs in an anarchic system. For example, if State A threatens to cut off exactly \$1 billion of trade as a punishment against State B for invading State C, but State C contains resources worth >\$1 billion, then State A is effectively inviting an invasion. This problem motivated the development of policies like “strategic ambiguity,” which involve the non-disclosure of explicit costs and the deliberate clouding of intentions. Strategic ambiguity makes it impossible for hostile actors to evaluate the costs of bad behavior, and is one of many policies that can curb calculated evil.

### 4.4 Conclusion

Egyptian Ratscrew has proven itself to be a rich game with complex strategies that rely not only on speed or reflexes, but also on calculation and economic decision making. In spite of the rigor that computational game theory provides, many aspects of this classic American pass-time continue to elude the comprehension of game theory. Although we succeeded in identifying risk slapping as a broadly dominant strategy, we remain captivated by the many unexpected implications of our research. In particular, we are captivated by the fact that after running 6.7 million simulations of the game, it feels as though we understand less about Egyptian Ratscrew than when we started.

It remains unclear to us what constitutes the perfect strategy – if such a strategy even exists. Nonetheless, we hope that in the future certain steps will be taken in order to improve our model and understand Egyptian Ratscrew more completely. For example, we ignored the implementation of mixed-strategies in our modeling, but hope that they will be comprehensively studied in the future. In game theory, as in all sciences, our knowledge and work is never complete.